# Quantum Mechanical View of Mathematical Statistics

Yu. I. Bogdanov
*OAO "Angstrem", Moscow, Russia*
E-mail: bogdanov@angstrem.ru

March 03, 2003

**Abstract**
Multiparametric statistical model providing stable reconstruction of parameters by observations is considered. The only general method of this kind is the root model based on the representation of the probability density as a squared absolute value of a certain function, which is referred to as a psi function in analogy with quantum mechanics. The psi function is represented by an expansion in terms of an orthonormal set of functions. It is shown that the introduction of the psi function allows one to represent the Fisher information matrix as well as statistical properties of the estimator of the state vector (state estimator) in simple analytical forms. A new statistical characteristic, a confidence cone, is introduced instead of a standard confidence interval. The chi-square test is considered to test the hypotheses that the estimated vector converges to the state vector of a general population and that both samples are homogeneous. The expansion coefficients are estimated by the maximum likelihood method. An iteration algorithm for solving the likelihood equation is presented. The stability and rate of convergence of the solution are studied. A special iteration parameter is introduced: its optimal value is chosen on the basis of the maximin strategy. Numerical simulation is performed using the set of the Chebyshev—Hermite functions as a basis. The problem of choosing an optimal number of harmonics in the expansion is discussed.

**Introduction**

As is well known, a basic quantum mechanical object completely determining the state of a quantum system is the psi function. Depending on a problem under consideration, other terms equivalent to the psi function, i.e., wave function, state vector, and probability amplitude are used. The psi function was introduced by Louis de Broglie and Ervin Shrödinger in wave mechanics. A probability interpretation implying that the square of the absolute value of the psi function describes the probability or probability density was proposed by Max Born.

All over the history of quantum physics, the concept of psi function as a basic object of quantum theory has been severely criticized. The well known examples are the theory of the pilot wave (Louis de Broglie [1]), the Bohm attempt to interpret quantum mechanics on the basis of hidden parameters [2], the Einstein doubts about the completeness of quantum mechanics [3], and his famous discussion with Bohr [4].

An idea to eliminate the psi function in the theory has seemed very attractive to many authors so far. Psychological reason for this phenomenon is apparently in the fact that classical (pre-quantum) theoretical probabilistic notions based on observable probabilities and probability densities seem to be clear and evident. At the same time, new quantum approach based on unobservable state vectors in complex Hilbert space appears too abstract and unnatural. That is seemingly why multiple attempts are made to refute standard interpretation of quantum mechanics and substitute it for a "causal" interpretation based on classical notions of probability and statistical laws.

Quantum mechanics is evidently statistical theory. On the other hand, its mathematical tools have little in common with such conventionally statistical scientific branches as probability theory, mathematical statistics, theory of random processes, and classical statistical physics. This fact noted even at the beginning of quantum theory has been a reason for numerous attempts, figuratively speaking, to bring a "lost quantum lamb" back to the general statistical "flock". As is well known, these attempts have failed so far (however, considering and criticizing these works are beyond the purposes of this paper). We do not also intend to propose a new interpretation of quantum mechanics as well as developing any axiomatic basis of either quantum mechanics or its formal mathematical tools [5-8]. Moreover, the initial statement of our problem had no concern with quantum theory, but was related to statistical data analysis [9,10]. Our aim was to develop a multiparametric statistical model providing stable reconstruction of unknown parameters by experimental data (see below for more detail). It turns out that there is actually only one statistical model of that kind, i.e., the model considered in quantum mechanics. In view of such results, the quantum mechanical model has essential advances compared to other statistical models. In other words, this model is special. Therefore, one cannot exclude that a "lost quantum lamb" may, in the course of time, lead all the statistical "flock".

We will consider the problem of probabilistic laws from the pragmatic applied point of view rather than from either philosophical or formally mathematical. Let us proceed from the fact that the subjects of probability theory and mathematical statistics are the direct and inverse statistical problems, respectively. Solving an inverse problem is essentially more complex problem than solving the direct one. Any statement (specification) of a direct problem by far not makes it possible to successfully solve the corresponding inverse one. The direct problem statement implies presetting the structures of probabilities and probability densities appearing in the mathematical model of a physical phenomenon. To solve an inverse problem, in its turn, means to find a constructive technique to estimate unknown probability structure and probability density on the basis of experimental observations.

From pragmatic point of view, it is purposeful to specify a direct problem so that there would exist a constructive and, wherever possible, simple solution of inverse problem. In fact, it turns out that the only way to do this is to go from the probability density to the psi function (root density estimator). Thus, the psi function becomes a mathematical object of data analysis that is introduced to provide effective constructive solving the inverse statistical problem. Validation and development of this statement is a basic aim of this paper.

In present paper, we try to show that a conventional formalism of mathematical statistics, which does not use the psi function, is not so attractive as it seems prima facie. Indeed, classical objects of mathematical statistics, as a rule, are smooth parameterized families of densities with either one or two unknown parameters to be estimated using the observed data. The functional form of the density is assumed to be initially prescribed. Such a parametric analysis is well developed to estimate the parameters of a rather small number of distributions (including Gaussian, exponential, binomial, Poisson and several other specific distributions). The basic limitation of the traditional parametric approach is that it is impossible to describe distributions of an arbitrary form. Attempts to estimate more complicated distributions, e.g., involving 3-4 or more unknown parameters run into severe calculation problems. In view of the ill-posedness of corresponding inverse problems, the numerical difficulties as well as those related to algorithm instabilities rapidly become insuperable with increasing dimension of a problem. It turns out that the only model providing stable consistent solutions in multidimensional problems is the model based on the concept of psi function.

Thus, the language of probability densities is not constructive in solving problems of statistical data analysis. This is some kind of a one way road. A statistical model has to be initially (a priory) given in this approach. Then, statistical properties of a system that can be further verified by an experiment can be found using the formalism of the probability theory and statistical physics. In practice, the situation is quite the contrary: we have some observations, whereas unknown state of the system has to be estimated on these data. However this is that kind of problems that mathematical statistics in its present form is unable to solve. Traditional mathematical statistics without the psi function looks like blindfold sack-race. Statisticians ignore solving the inverse statistical problem, which is a fundamental problem in statistics, substituting it for solving more specific particular problems. Besides that, their activities are confined within the narrow bounds of models with extremely limited number of degrees of freedom that are at their disposal.

In our opinion, it is remarkable that the basic object of quantum theory (psi function) is necessary to effectively solve the inverse statistical problem. Indeed, quantum mechanics describes phenomena where statistical laws manifest themselves in most evident and fundamental form (in contrast to classical statistics historically originated from problems related to gambling, and allowed Einstein to draw an analogy between statistical nature of quantum mechanics and notion of Nature "playing dice"). If Nature uses any other language than that of the psi function (e.g., "plays dice" by classical random processes), we would have extremely limited opportunities to interpret its "messages". In other words, we could not in fact reconstruct the structure of states under study by the experimental data, since the corresponding mathematical problem would be ill-posed. Thus, Nature chose the language of psi function, for we can understand only this language (cognizableness). Only the language of quantum mechanics makes it possible to describe the structure of complex systems and their interactions in a simple form disregarding an infinite number of insignificant details. Only this language makes the dialogue between man and Nature possible. Owing to this language, man can not only describe the structure of complex states, e.g., such as those describing polyatomic molecules and crystals, but has also an opportunity to prepare these states and use their dynamics for his own purposes. This has already done, for instance, in lasers, micro- and nanoelectronic devices, and is going to be used in, e.g., quantum computer. In other words, "quantum physics is an elementary theory of information" [11].

This paper is based on a symbiosis of mathematical tools of quantum mechanics and the Fisher maximum likelihood principle (Appendix 1) in order to find effective state estimators with most simple (and fundamental) statistical properties.

The method proposed, the root density estimator, is based on the representation of the density in the form of a squared absolute value of a certain function, which is referred to as a psi function in analogy with quantum mechanics [9,10].

The introduction of the psi function results in substantial reduction in the structure of both the Fisher information matrix and covariance matrix of estimators, making them independent of a basis, allows one to provide positive definiteness of the density and represent results in a simple analytical form.

The root density estimator being based on the maximum likelihood method has optimal asymptotic behavior in contrast to kernel and orthogonal series estimators.

The likelihood equation in the method of the root density estimator has a simple quasilinear structure and admits developing effective rapidly converging iteration procedure even in the case of multiparametric problems (for example, when the number or parameters to be estimated runs up to many tens or even hundreds). That is why the problem under consideration favorably differs from the other well-known problems solved by the maximum likelihood method when the complexity of numerical simulations rapidly increases and the stability of algorithms decreases with increasing number of parameters to be estimated.

Basic objects of the theory (state vectors, information and covariance matrices etc.) become simple geometrical objects in the Hilbert space that are invariant with respect to unitary (orthogonal) transformations.

We are considering the application of this method to the problems of estimating the quantum states in the next paper ("Root Estimator of Quantum States ").

## 1. Fisher Information Matrix and Density Estimator

A psi function considered further is a mathematical object of statistical data analysis. This function is introduced in the same way as in quantum mechanics to drastically simplify statistical state estimators. The introduction of the psi function implies that the "square root" of the density distribution function is considered instead of the density distribution function itself.

$$p(x) = |\psi(x)|^2 \quad (1.1)$$

Let the psi function depend on $s$ unknown parameters $c_0, c_1, ..., c_{s-1}$ (according to quantum mechanics, the basis functions are traditionally numbered from zero corresponding to the ground state). The parameters introduced are the coefficients of an expansion in terms of a set of basis functions:

$$\psi(x) = \sum_{i=0}^{s-1} c_i \varphi_i(x). \quad (1.2)$$

Assume that the set of the functions is orthonormal. Then, the normalization condition (the total probability is equal to unity) is given by

$$c_i c_i^* = 1. \quad (1.3)$$

Hereafter, we imply the summation over recurring indices numbering the terms of the expansion in terms of basis functions (unless otherwise stated). On the contrary, statistical sums denoting the summation over the sample points will be written in an explicit form.

For the sake of simplicity, consider first a real valued psi function.

Let an expansion have the form

$$\psi(x) = \sqrt{1 - (c_1^2 + \ldots + c_{s-1}^2)}\varphi_0(x) + c_1\varphi_1(x) + \ldots + c_{s-1}\varphi_{s-1}(x). \quad (1.4)$$

Here, we have eliminated the coefficient $c_0 = \sqrt{1 - (c_1^2 + \ldots + c_{s-1}^2)}$ from the set of parameters to be estimated, since it is expressed via the other coefficients by the normalization condition.

The parameters $c_1, c_2, \ldots, c_{s-1}$ are independent. We will study their asymptotic behavior using the Fisher information matrix [12-14]

$$I_{ij}(c) = n \cdot \int \frac{\partial \ln p(x,c)}{\partial c_i} \frac{\partial \ln p(x,c)}{\partial c_j} p(x,c) dx. \quad (1.5)$$

For analogy between the Fisher information and action functional, see [15]. In problems of estimating quantum states, the Fisher information matrix was particularly considered in [16,17].

It is of particular importance for our study that the Fisher information matrix drastically simplifies if the psi function is introduced:

$$I_{ij} = 4n \cdot \int \frac{\partial \psi(x,c)}{\partial c_i} \frac{\partial \psi(x,c)}{\partial c_j} dx. \quad (1.6)$$

In the case of the expansion (1.4), the information matrix $I_{ij}$ is $(s-1) \times (s-1)$ matrix of the form:

$$I_{ij} = 4n\left(\delta_{ij} + \frac{c_i c_j}{c_0^2}\right), \quad c_0 = \sqrt{1 - (c_1^2 + \ldots + c_{s-1}^2)}. \quad (1.7)$$

A noticeable feature of the expression (1.7) is its independence on the choice of basis functions. Let us show that only the root density estimator has this property.

Consider the following problem that can be referred to as a generalized orthogonal series density estimator. Let the density $p$ be estimated by a composite function of another (for simplicity, real-valued) function $g$. The latter function, in its turn, is represented in the form of the expansion in terms of a set of orthonormal functions, i.e.,

$$p = p(g), \text{ where } g(x) = \sum_{i=0}^{s-1} c_i \varphi_i(x). \quad (1.8)$$

We assume that the norm of the function $g$ is finite:

$$N_g^2 = \|g\|^2 = \int g^2(x) dx = \sum_{i=0}^{s-1} c_i^2 \quad (1.9)$$

Let the coefficients $c_i$; $i = 0, 1, \ldots, s-1$ be estimated by the maximum likelihood method.

Consider the following matrix:

$$\tilde{I}_{ij}(c) = n \cdot \int \frac{\partial \ln p(x,c)}{\partial c_i} \frac{\partial \ln p(x,c)}{\partial c_j} p(x,c) dx =$$

$$= n\int \frac{1}{p} \frac{\partial p(x,c)}{\partial c_i} \frac{\partial p(x,c)}{\partial c_j} dx = n\int \frac{1}{p}\left(\frac{\partial p}{\partial g}\right)^2 \frac{\partial g(x,c)}{\partial c_i} \frac{\partial g(x,c)}{\partial c_j} dx \quad (1.10)$$

The structure of this matrix is simplest if its elements are independent of both the density and basis functions. This can be achieved if (and only if) $p(g)$ satisfies the condition

$$\frac{1}{p}\left(\frac{\partial p}{\partial g}\right)^2 = const$$

yielding

$$g = const\sqrt{p} \quad (1.11)$$

Choosing unity as the constant in the last expression, we arrive at the psi function $g = \psi = \sqrt{p}$ with the simplest normalization condition (1.3).

The $\tilde{I}$ matrix has the form

$$\tilde{I}_{ij} = 4n\delta_{ij} \qquad i,j = 0,1,\ldots,s-1 \quad (1.12)$$

The $\tilde{I}$ matrix under consideration is not the true Fisher information matrix, since the expansion parameters $c_i$ are dependent. They are related to each other by the normalization condition. That is why we will refer to this matrix as a prototype of the Fisher information matrix.

As is seen from the asymptotic expansion of the log likelihood function in the vicinity of a stationary point, statistical properties of the distribution parameters are determined by the quadratic form $\sum_{i,j=0}^{s-1}\tilde{I}_{ij}\delta c_i \delta c_j$. Separating out zero component of the variation and taking into account that

$$\delta c_0^2 = \sum_{i,j=1}^{s-1} \frac{c_i c_j}{c_0^2} \delta c_i \delta c_j \quad \text{(see the expression (1.16) below), we find}$$

$$\sum_{i,j=0}^{s-1}\tilde{I}_{ij}\delta c_i \delta c_j = \sum_{i,j=1}^{s-1} I_{ij}\delta c_i \delta c_j \quad , (1.13)$$

where the true information matrix $I$ has the form of (1.7).

Thus, the representation of the density in the form $p = |\psi|^2$ (and only this representation) results in a universal (and simplest) structure of the Fisher information matrix.

In view of the asymptotic efficiency, the covariance matrix of the state estimator is the inverse Fisher information matrix:

$$\Sigma(\hat{c}) = I^{-1}(c) \quad (1.14)$$

The matrix components are

$$\Sigma_{ij} = \frac{1}{4n}(\delta_{ij} - c_i c_j) \qquad i,j = 1,\ldots,s-1 \quad (1.15)$$

Now, let us extend the covariance matrix found by appending the covariance between the $c_0$ component of the state vector and the other components.

Note that

$$\delta c_0 = \frac{\partial c_0}{\partial c_i} \delta c_i = \frac{-c_i}{c_0} \delta c_i. \quad (1.16)$$

This yields

$$\Sigma_{0j} = \overline{\delta c_0 \delta c_j} = \frac{-c_i}{c_0} \overline{\delta c_i \delta c_j} =$$

$$= \frac{-\Sigma_{ji} c_i}{c_0} = \frac{-c_i (\delta_{ji} - c_j c_i)}{4 n c_0} = -\frac{c_0 c_j}{4n}. \quad (1.17)$$

Similarly,

$$\Sigma_{00} = \overline{\delta c_0 \delta c_0} = \frac{c_i c_j}{c_0^2} \overline{\delta c_i \delta c_j} = \frac{c_i c_j}{c_0^2} \Sigma_{ij} = \frac{1 - c_0^2}{4n}. \quad (1.18)$$

Finally, we find that the covariance matrix has the same form as (1.15):

$$\Sigma_{ij} = \frac{1}{4n} (\delta_{ij} - c_i c_j) \quad i, j = 0, 1, \ldots, s-1. \quad (1.19)$$

This result seems to be almost evident, since the zero component is not singled out from the others (or more precisely, it has been singled out to provide the fulfillment of the normalization condition). From the geometrical standpoint, the covariance matrix (1.19) is a second-order tensor.

Moreover, the covariance matrix (up to a constant factor) is a single second-order tensor satisfying the normalization condition.

Indeed, according to the normalization condition,

$$\delta (c_i c_i) = 2 c_i \delta c_i = 0. \quad (1.20)$$

Multiplying the last equation by an arbitrary variation $\delta c_j$ and averaging over the statistical ensemble, we find

$$c_i E(\delta c_i \delta c_j) = \Sigma_{ji} c_i = 0. \quad (1.21)$$

Only two different second-order tensors can be constructed on the basis of the vector $c_i$: $\delta_{ij}$ and $c_i c_j$. In order to provide the fulfillment of (1.21) following from the normalization condition, these tensors have to appear in the matrix only in the combination (1.19).

It is useful to consider another derivation of the covariance matrix. According to the normalization condition, the variations $\delta c_i$ are dependent, since they are related to each other by the linear relationship (1.20). In order to make the analysis symmetric (in particular, to avoid expressing one component via the others as it has been done in (1.4)), one may turn to other variables that will be referred to as principle components.

Consider the following unitary (orthogonal) transformation:

$$U_{ij} \delta c_j = \delta f_i \qquad i, j = 0, 1, \ldots, s-1. \quad (1.22)$$

Let the first (to be more precise, zero) row of the transformation matrix coincide with the state vector: $U_{0j} = c_j$. Then, according to (1.20), the zero variation is identically zero in new coordinates: $\delta f_0 = 0$.

The inverse transformation is

$$U_{ij}^{+} \delta f_j = \delta c_i \qquad i, j = 0, 1, \ldots, s-1. \quad (1.23)$$

In view of the fact that $\delta f_0 = 0$, the first (more precisely, zero) column of the matrix $U^+$ can be eliminated turning the matrix into the factor loadings matrix $L$. Then

$$\delta c_i = L_{ij}\delta f_j \qquad i = 0,1,...,s-1; \quad j = 1,...,s-1. \quad (1.24)$$

The relationship found shows that $s$ components of the state-vector variation are expressed through $s-1$ principal components (that are independent Gaussian variables).

In terms of principle components, the Fisher information matrix and covariance matrix are proportional to a unit matrix:

$$I^f_{ij} = 4n\delta_{ij} \qquad i,j = 1,...,s-1, \quad (1.25)$$

$$\Sigma^f_{ij} = \overline{\delta f_i \delta f_j} = \frac{\delta_{ij}}{4n} \qquad i,j = 1,...,s-1. \quad (1.26)$$

The last relationship particularly shows that the principal variation components are independent and have the same variance $\frac{1}{4n}$.

The expression for the covariance matrix of the state vector components can be easily found on the basis of (1.26). Indeed,

$$\Sigma_{ij} = \overline{\delta c_i \delta c_j} = L_{ik}L_{js}\overline{\delta f_k \delta f_s} = L_{ik}L_{js}\frac{\delta_{ks}}{4n} = \frac{L_{ik}L_{jk}}{4n}. \quad (1.27)$$

In view of the unitarity of the $U^+$ matrix, we have

$$L_{ik}L_{jk} + c_i c_j = \delta_{ij}. \quad (1.28)$$

Taking into account two last formulas, we finally find the result presented above:

$$\Sigma_{ij} = \frac{1}{4n}(\delta_{ij} - c_i c_j) \qquad i,j = 0,1,...,s-1. \quad (1.29)$$

In quantum mechanics, the matrix

$$\rho_{ij} = c_i c_j \quad (1.30)$$

is referred to as a density matrix (of a pure state). Thus,

$$\Sigma = \frac{1}{4n}(E - \rho), \quad (1.31)$$

where $E$ is the $s \times s$ unit matrix.

In the diagonal representation,

$$\Sigma = UDU^+, \quad (1.32)$$

where $U$ and $D$ are unitary (orthogonal) and diagonal matrices, respectively.

As is well known from quantum mechanics and readily seen straightforwardly, the density matrix of a pure state has the only (equal to unity) element in the diagonal representation. Thus, in our case, the diagonal of the $D$ matrix has the only element equal to zero (the corresponding eigenvector is the state vector); whereas the other diagonal elements are equal to $\frac{1}{4n}$ (corresponding eigenvectors and their linear combinations form a subspace that is orthogonal complement to the state vector). The zero element at a principle diagonal indicates that the inverse matrix (namely, the Fisher information matrix of the $s$-th order) does not exist. It is clear since there are only $s-1$ independent parameters in the distribution.

The results on statistical properties of the state vector reconstructed by the maximum likelihood method can be summarized as follows. In contrast to a true state vector, the estimated one involves noise in the form of a random deviation vector located in the space orthogonal to the true state vector. The components of the deviation vector (totally, $s-1$ components) are asymptotically normal independent random variables with the same variance $\frac{1}{4n}$. In the aforementioned $s-1$-dimensional space, the deviation vector has an isotropic distribution, and its squared length is the random variable $\frac{\chi^2_{s-1}}{4n}$, where $\chi^2_{s-1}$ is the random variable with the chi-square distribution of $s-1$ degrees of freedom, i.e.,

$$c_i = (c, c^{(0)}) \cdot c_i^{(0)} + \xi_i \qquad i = 0,1,...,s-1. \quad (1.33)$$

where $c^{(0)}$ and $c$ are true and estimated state vectors, respectively; $(c, c^{(0)}) = c_i c_i^{(0)}$, their scalar product; and $\xi_i$, the deviation vector. The deviation vector is orthogonal to the vector $c^{(0)}$ and has the squared length of $\frac{\chi^2_{s-1}}{4n}$ determined by chi-square distribution of $s-1$ degrees of freedom, i.e.,

$$(\xi, c^{(0)}) = \xi_i c_i^{(0)} = 0 \quad (\xi, \xi) = \xi_i \xi_i = \frac{\chi^2_{s-1}}{4n} \qquad (1.34)$$

Squaring (1.33), in view of (1.34), we have

$$1 - (c, c^{(0)})^2 = \frac{\chi^2_{s-1}}{4n}. \quad (1.35)$$

This expression means that the squared scalar product of the true and estimated state vectors is smaller than unity by asymptotically small random variable $\frac{\chi^2_{s-1}}{4n}$.

The results found allow one to introduce a new stochastic characteristic, namely, a confidence cone (instead of a standard confidence interval). Let $\vartheta$ be the angle between an unknown true state vector $c^{(0)}$ and that $c$ found by solving the likelihood equation. Then,

$$\sin^2 \vartheta = 1 - \cos^2 \vartheta = 1 - (c, c^{(0)})^2 = \frac{\chi^2_{s-1}}{4n} \leq \frac{\chi^2_{s-1,\alpha}}{4n}. \quad (1.36)$$

Here, $\chi^2_{s-1,\alpha}$ is the quantile corresponding to the significance level $\alpha$ for the chi-square distribution of $s-1$ degrees of freedom.

The set of directions determined by the inequality (1.36) constitutes the confidence cone. The axis of a confidence cone is the reconstructed state vector $c$. The confidence cone covers the direction of an unknown state vector at a given confidence level $P = 1 - \alpha$.

From the standpoint of theory of unitary transformations in quantum mechanics (in our case transformations are reduced to orthogonal), it can be found an expansion basis in (1.4) such that the sum will contain the only nonzero term, namely, the true psi function. This result means that if the best basis is guessed absolutely right and the true state vector is $(1,0,0,...,0)$, the empirical state vector estimated by the maximum likelihood method will be the random vector $(c_0, c_1, c_2, ..., c_{s-1})$, where $c_0 = \sqrt{1 - (c_1^2 + ... + c_{s-1}^2)}$, and the other components $c_i \sim N\left(0, \frac{1}{4n}\right)$ $i = 1,...,s-1$ will be independent as it has been noted earlier.

## 2. Chi-Square Criterion. Test of the Hypothesis That the Estimated State Vector Equals to the State Vector of a General Population. Estimation of the Statistical Significance of Differences between Two Samples.

Rewrite (1.35) in the form

$$4n\left(1-\left(c,c^{(0)}\right)^2\right)=\chi^2_{s-1}. \quad (2.1)$$

This relationship is a chi-square criterion to test the hypothesis that the state vector estimated by the maximum likelihood method $c$ equals to the state vector of general population $c^{(0)}$.

In view of the fact that for $n\to\infty$ $1+\left(c,c^{(0)}\right)\to 2$, the last inequality may be rewritten in another asymptotically equivalent form

$$4n\sum_{i=0}^{s-1}\left(c_i-c_i^{(0)}\right)^2=\chi^2_{s-1}. \quad (2.2)$$

Here, we have taken into account that

$$\sum_{i=0}^{s-1}\left(c_i-c_i^{(0)}\right)^2=\sum_{i=0}^{s-1}\left(c_i^2-2c_ic_i^{(0)}+c_i^{(0)2}\right)=\sum_{i=0}^{s-1}2\left(1-c_ic_i^{(0)}\right).$$

As is easily seen, the approach under consideration involves the standard chi-square criterion as a particular case corresponding to a histogram basis (see Sec.4). Indeed, the chi-square parameter is usually defined as [12]

$$\chi^2=\sum_{i=0}^{s-1}\frac{\left(n_i-n_i^{(0)}\right)^2}{n_i^{(0)}},$$

where $n_i^{(0)}$ is the number of points expected in $i$-th interval according to theoretical distribution.

In the histogram basis (see Sec.4 ), $c_i=\sqrt{\frac{n_i}{n}}$ is the empirical state vector and $c_i^{(0)}=\sqrt{\frac{n_i^{(0)}}{n}}$, theoretical state vector. Then,

$$\chi^2=n\sum_{i=0}^{s-1}\frac{\left(c_i^2-c_i^{(0)2}\right)^2}{c_i^{(0)2}}\to 4n\sum_{i=0}^{s-1}\left(c_i-c_i^{(0)}\right)^2. \quad (2.3)$$

Here, the sign of passage to the limit means that random variables appearing in both sides of (2.3) have the same distribution. We have used also the asymptotic approximation

$$c_i^2-c_i^{(0)2}=\left(c_i+c_i^{(0)}\right)\left(c_i-c_i^{(0)}\right)\to 2c_i^{(0)}\left(c_i-c_i^{(0)}\right). \quad (2.4)$$

Comparing (2.2) and (2.3) shows that the parameter $\chi^2$ is a random variable with $\chi^2$-distribution of $s-1$ degrees of freedom (if the tested hypothesis is valid). Thus, the standard chi-square criterion is a particular case (corresponding to a histogram basis) of the general approach developed here (that can be used in arbitrary basis).

The chi-square criterion can be applied to test the homogeneity of two different set of observations (samples). In this case, the hypothesis that the observations belong to the same statistical ensemble (the same general population) is tested.

In the case under consideration, the standard chi-square criterion [12] may be represented in new terms as follows:

$$\chi^2 = n_1 n_2 \sum_{i=0}^{s-1} \frac{\left(\frac{n_i^{(1)}}{n_1} - \frac{n_i^{(2)}}{n_2}\right)^2}{n_i^{(1)} + n_i^{(2)}} \to 4 \frac{n_1 n_2}{n_1 + n_2} \sum_{i=0}^{s-1} \left(c_i^{(1)} - c_i^{(2)}\right)^2, \quad (2.5)$$

where $n_1$ and $n_2$ are the sizes of the first and second samples, respectively; $n_i^{(1)}$ and $n_i^{(2)}$, the numbers of points in the $i$-th interval; and $c_i^{(1)}$ and $c_i^{(2)}$, the empirical state vectors of the samples. In the left side of (2.5), the chi-square criterion in a histogram basis is presented; in the right side, the same criterion in general case. The parameter $\chi^2$ defined in such a way is a random variable with the $\chi^2$ distribution of $s-1$ degrees of freedom (the sample homogeneity is assumed).

The chi-square criterion (2.5) can be represented in the form similar to (2.1) as a measure of proximity to unity of the absolute value of scalar product of sample state vectors:

$$4 \frac{n_1 n_2}{n_1 + n_2} \left(1 - \left(c^{(1)} c^{(2)}\right)^2\right) = \chi^2_{s-1} \qquad (2.6)$$

### 3. Psi Function and Likelihood Equation

The maximum likelihood method is the most popular classical method for estimating the parameters of the distribution density [12-14]. In problems of estimating quantum states, this method was first used in [18,19].

We will consider sets of basis functions that are complete for $s \to \infty$. At a finite $s$, the estimation of the function by (1.2) involves certain error. The necessity to restrict the consideration to a finite number of terms is related to the ill-posedness of the problem [20]. Because of the finite set of experimental data, a class of functions, where the psi function is sought, should not be too wide; otherwise a stable description of a statistical distribution would be impossible. Limiting the number of terms in the expansion by a finite number $s$ results in narrowing the class of functions where a solution is sought. On the other hand, if the class of functions turns out to be too narrow (at too small $s$), the dependence to be found would be estimated within too much error. The problem of an optimal choice of the number of terms in the expansion is discussed in greater detail in Sec.6.

The maximum likelihood method implies that the values maximizing the likelihood function and its logarithm

$$\ln L = \sum_{k=1}^{n} \ln p(x_k | c) \to \max \qquad (3.1)$$

are used as most likely estimators for unknown parameters $c_0, c_1, \ldots, c_{s-1}$.

The probability density is

$$p(x) = \psi^* \psi = c_i c_j^* \varphi_i(x) \varphi_j^*(x). \qquad (3.2)$$

At sample points, the distribution density is

$$p(x_k) = \psi^* \psi = c_i c_j^* \varphi_i(x_k) \varphi_j^*(x_k). \qquad (3.3)$$

In our case, the likelihood function has the form

$$\ln L = \sum_{k=1}^{n} \ln\left[c_i c_j^* \varphi_i(x_k) \varphi_j^*(x_k)\right]. \quad (3.4)$$

In view of the normalization condition, seeking an extremum of the log likelihood function is reduced to that for the following function:

$$S = \ln L - \lambda\left(c_i c_i^* - 1\right), \quad (3.5)$$

where $\lambda$ is the Lagrange multiplier.

The necessary condition for an extremum yields the likelihood equation

$$\frac{\partial S}{\partial c_i^*} = \sum_{k=1}^{n} \frac{\varphi_i^*(x_k)\varphi_j(x_k)}{p(x_k)} c_j - \lambda c_i = 0. \quad (3.6)$$

Thus, the problem of looking for the extremum is reduced to the eigenvalue problem

$$R_{ij} c_j = \lambda c_i \qquad i, j = 0, 1, \ldots, s - 1, \quad (3.7)$$

where

$$R_{ij} = \sum_{k=1}^{n} \frac{\varphi_i^*(x_k)\varphi_j(x_k)}{p(x_k)}. \quad (3.8)$$

The problem (3.7) is formally linear. However, the matrix $R_{ij}$ depends on an unknown density $p(x)$. Therefore, the problem under consideration is actually nonlinear, and should be solved by the iteration method (see Sec.5).

An exception is the histogram density estimator presented below when the problem can be solved straightforwardly (see Sec.4). Multiplying both parts of Eq. (3.7) by $c_i^*$ and summing with respect to $i$, in view of (1.3) and (3.2), we find that the most likely state vector $c$ always corresponds to its eigenvalue $\lambda = n$.

Let us verify whether the substitution of the true state vector into the likelihood equation turns it into an identical relation (in the asymptotic limit). Indeed, at a large sample size ($n \to \infty$), according to the law of large numbers (the sample mean tends to the population mean) and the orthonormality of basis functions, we have

$$\frac{1}{n} R_{ij} = \frac{1}{n} \sum_{k=1}^{n} \frac{\varphi_i^*(x_k)\varphi_j(x_k)}{p(x_k)} \to$$

$$\to \int \frac{\varphi_i^*(x)\varphi_j(x)}{p(x)} p(x) dx = \delta_{ij} \quad (3.9)$$

Thus, the matrix $\frac{1}{n}R$ asymptotically tends to unit matrix. In other words, Eq. (3.7) shows that the true state vector is its solution for $n \to \infty$ (consistency). The matrix $\frac{1}{n}R$ may be referred to as a quasi-unit.

Let us assume that the basis functions $\varphi_i(x)$ and the state vector $c$ are real valued. Then, the basic equation for the state vector (3.7) can be expressed in the form

$$\frac{1}{n}\sum_{k=1}^{n}\left(\frac{\varphi_i(x_k)}{\sum_{j=0}^{s-1}c_j\varphi_j(x_k)}\right) = c_i \qquad i = 0,1,\ldots,s-1 \quad . \quad (3.10)$$

Here, we have written the summation signs for clearness. As is easily seen, the solution of this equation satisfies the normalization condition (1.3).

4. **Histogram Density Estimator**

In order to study the histogram density estimator, one has to assume that a distribution is given in a finite region (in the case of variables distributed along an infinite interval, it is necessary to cut off the distribution tails, e.g., by using maximum and minimum values in the sample as bounds).

Let us divide the full range of variation for a random variable into a finite number of intervals. Points $x_0, x_1, \ldots, x_s$ divide the full range of variation for a random variable into $s$ intervals (bins).

Assume that

$$\varphi_i(x) = \begin{cases} \dfrac{1}{(x_{i+1}-x_i)^{1/2}} & \text{at} \quad x_i \le x \le x_{i+1} \\ 0 & \text{otherwise} \end{cases} \quad . \quad (4.1)$$

The functions $\varphi_i(x)$ $i = 0,1,\ldots,s-1$ form an orthonormal but, of course, incomplete set.

Equation (3.10) yields the following most likely estimator for the psi function:

$$\psi(x) = \sum_{i=0}^{s-1} c_i \varphi_i(x), \qquad c_i = \sqrt{\frac{n_i}{n}}, \quad (4.2)$$

where $n_i$ is the number of points in $i$-th interval.

In order to avoid appearing indeterminate forms (zero divided by zero) while calculating the expansion coefficients $c_i$, one has to assume that $n_i > 0$ in each interval.

As is easily seen, the square of the psi function constructed in this way is a histogram density estimator.

Note, that the root transformation of a histogram is used in statistics to find another graphic presentation of data that is referred to as a rootgram [21]. In contrast to histogram, the rootgram has a uniform variance.

Applying a unitary transformation to the found state vector shows a natural way to smooth a histogram density estimator that is as follows.

Let us transform the column vector $c_i$ and basis functions $\varphi_i(x)$ by a unitary matrix $U$:

$$c'_i = U_{ik} c_k , \quad (4.3)$$

$$\varphi'_i(x) = U^*_{il} \varphi_l(x) . \quad (4.4)$$

The psi function and, hence, the density turn out to be invariant with respect to this transformation. Indeed,

$$\psi'(x) = c_i' \varphi_i'(x) = U_{ik} c_k U_{il}^* \varphi_l(x) = c_l \varphi_l(x). \quad (4.5)$$

Here, we have taken into account that due to the unitarity of the matrix $U$,

$$U_{il}^* U_{ik} = U_{li}^+ U_{ik} = \delta_{lk}. \quad (4.6)$$

The aforementioned transformation will be useful if the basis functions $\varphi_i'(x)$ in a new representation may be ranged in increasing complexity in such a way that the amplitudes $c_i'$ corresponding to first (most simple) basis functions turn out to be large, whereas those corresponding to more complex basis functions, relatively small. Then, a histogram density can be smoothed by truncating higher harmonics.

A classical example of such a unitary transformation is the discrete Fourier transform given by the matrix

$$U_{kl} = \frac{1}{\sqrt{s}} \exp\left(i \frac{2\pi}{s} kl\right). \quad (4.7)$$

A state vector resulting from the unitary transformation with the matrix (4.7) may be interpreted as a frequency spectrum of a signal that is the histogram estimator of a psi function.

In general case, choosing a unitary transformation and a way in which to filter noise, and ranging basis functions in ascending order of complexity should be performed on the basis of the analysis of a particular problem.

The theoretical state vector in the histogram basis is

$$c_i^0 = \sqrt{p_i} \quad i = 1,...,s \quad (4.8)$$

Here, $p_1, p_2, ..., p_s$ are the theoretical probabilities of finding the point in the corresponding interval.

In view of (4.2), the chi-square criterion (2.1) applied to the histogram basis yields

$$4\left[n - \left(\sqrt{n_1 p_1} + \sqrt{n_2 p_2} + ... + \sqrt{n_s p_s}\right)^2\right] = \chi_{s-1}^2. \quad (4.9)$$

Here, $n_1, n_2, ..., n_s$ are number of points observed in the corresponding intervals.

The expression (4.9) implies that if the zero hypothesis is valid, i.e., the probability distribution corresponds to the theoretical one, the left side is a random variable with the chi-square distribution of $s - 1$ degrees of freedom.

The standard form of the chi-square criterion follows from the chi-square criterion in the form (4.9) in the asymptotical limit.

Let us consider in detail the case of $s = 2$ corresponding to the binomial distribution. The exact state vector $c^0$ and its estimation $c$ are

$$c^0 = \begin{pmatrix} \sqrt{p_1} \\ \sqrt{p_2} \end{pmatrix} \quad c = \frac{1}{\sqrt{n}} \begin{pmatrix} \sqrt{n_1} \\ \sqrt{n_2} \end{pmatrix} \quad (4.10)$$

Consider the transformation of these vectors by the following unitary (orthogonal) matrix:

$$U = \begin{pmatrix} \sqrt{p_1} & \sqrt{p_2} \\ \sqrt{p_2} & -\sqrt{p_1} \end{pmatrix} \quad (4.11)$$

This yields

$$Uc^0 = \begin{pmatrix} 1 \\ 0 \end{pmatrix}, \qquad (4.12)$$

$$Uc = \frac{1}{\sqrt{n}} \begin{pmatrix} \sqrt{p_1 n_1} + \sqrt{p_2 n_2} \\ \sqrt{p_2 n_1} - \sqrt{p_1 n_2} \end{pmatrix} = \begin{pmatrix} \sqrt{1-\xi^2} \\ \xi \end{pmatrix}, \qquad (4.13)$$

where $\xi \sim N\left(0, \dfrac{1}{4n}\right)$

In the latter equality we have used the remark made in the end of Sec. 1. We represent the second row of the relationship in the form

$$2\left(\sqrt{n_1 p_2} - \sqrt{n_2 p_1}\right) \sim N(0,1) \qquad (4.14)$$

Here, $p_1 + p_2 = 1$, $n_1 + n_2 = n$.

The relationship (4.14) describes the approximation of the binomial distribution by a normal one. Similar result of classical probability theory (Moivre-Laplace theorem) is [12]

$$\frac{n_1 - np_1}{\sqrt{np_1 p_2}} \sim N(0,1) \qquad (4.15)$$

It is easily seen that the formula (4.15) follows from (4.14) in the asymptotical limit. At the same time, the formula (4.14) is a better approximation in the case of a finite sample sizes than the classical result (4.15).

## 5. Computational Approach to Solving Likelihood Equation

In order to develop an iteration procedure for Eq. (3.10), let us rewrite it in the form

$$\alpha c_i + (1-\alpha) \frac{1}{n} \sum_{k=1}^{n} \left( \frac{\varphi_i(x_k)}{\sum_{j=0}^{s-1} c_j \varphi_j(x_k)} \right) = c_i . \qquad (5.1)$$

Here, we have introduced an additional parameter $0 < \alpha < 1$ that does not change the equation itself but substantially influences the solution stability and the rate of convergence of an iteration procedure.

Let us represent an iteration procedure (transition from $r$-th to $r+1$-th approximation) in the form

$$c_i^{r+1} = \alpha \cdot c_i^r + (1-\alpha) \cdot \frac{1}{n} \sum_{k=1}^{n} \left( \frac{\varphi_i(x_k)}{\sum_{j=0}^{s-1} c_j^r \varphi_j(x_k)} \right) . \qquad (5.2)$$

Let us study the conditions of stable convergence of the iteration procedure to the solution. We restrict our consideration to the case of small deviations. Let $\delta c$ be any small deviation of an approximate state vector from the exact solution of Eq. (5.1) at arbitrary step; and $\delta c'$, that at the next iteration step. The squared distance between the exact and approximate solutions is $(\delta c)^T (\delta c)$.

A fundamental condition for an iteration procedure to converge is that the corresponding mapping has to be contracting (see the principle of contracting mappings and fixed point theorem [22]). A mapping is contracting if $(\delta c')^T (\delta c') < (\delta c)^T (\delta c)$. It can be proved that

$$\delta c' = A \delta c, \quad (5.3)$$

where $A$ is the perturbation matrix:

$$A = \alpha E - \frac{(1-\alpha)}{n} R. \quad (5.4)$$

Here, $E$ is an $(s \times s)$ unit matrix.

After an iteration, the squared distance is decreased by

$$(\delta c)^T (\delta c) - (\delta c')^T (\delta c') = (\delta c)^T B (\delta c), \quad (5.5)$$

where

$$B = E - A^T A \quad (5.6)$$

is the contracting matrix.

The mapping is contracting if $B$ is a positive matrix.

The minimum eigenvalue $\lambda_{min}$ of the $B$ matrix is expedient to consider as a measure of contractility. An eigenfunction related to $\lambda_{min}$ corresponds to perturbation that is worst from the convergence standpoint. Thus, the parameter $\lambda_{min}$ characterizes the guaranteed convergence, since the squared distance decreases at least by $\lambda_{min} \cdot 100\%$ at each step.

Let $R_0$ be the vector of eigenvalues for the $R$ matrix.

The $B$-matrix eigenvalues are expressed in terms of the $R$-matrix eigenvalues by

$$\lambda_i = 1 - \alpha^2 + \frac{2\alpha(1-\alpha)}{n} R_{0i} - \frac{(1-\alpha)^2}{n^2} R_{0i}^2 \quad (5.7)$$

The minimum of this expression at any given $\alpha$ is determined by either maximum (at small $\alpha$) or minimum (at large $\alpha$) $R_{0i}$.

As an optimal value of $\alpha$, we will use the value at which $\lambda_{min}$ reaches its maximum (maximin rule). An optimal value of $\alpha$ is determined by the sum of maximum and minimum values of $R_{0i}$:

$$\alpha_{opt} = \frac{D_0}{2n + D_0}, \quad D_0 = \max(R_{0i}) + \min(R_{0i}). \quad (5.8)$$

For the difference of distances between the approximate and exact solutions before and after iteration, we have

$$\rho' \leq \varepsilon \rho, \text{ where } \varepsilon = \sqrt{1 - \lambda_{min}}. \quad (5.9)$$

The distance between the approximate $c^{(r)}$ and exact $c$ solutions decreases not slower than infinitely decreasing geometric progression.

$$\rho(c^{(r)}, c) \leq \varepsilon^r \rho(c^{(0)}, c) \leq \frac{\varepsilon^r}{1 - \varepsilon} \rho(c^{(0)}, c^{(1)}). \quad (5.10)$$

The result obtained implies that the number of iterations required for the distance between the approximate and exact solutions to decrease by the factor of $\exp(k_0)$ is

$$r_0 \approx \frac{-2k_0}{\ln(1-\lambda_{min})}. \quad (5.11)$$

As it follows from the above analysis and numerical simulations, the iteration procedure turns out to be unstable at small values of the iteration parameter $\alpha$ when the minimum eigenvalue $\lambda_{min}$ of the contracting matrix $B$ is negative ($\lambda_{min} < 0$). The instability arises as a result of the fact that the maximum eigenvalue of the matrix $R$ exceeds the sample size ($\max(R_{0i}) > n$). The critical value $\alpha_c$ of the iteration parameter corresponding to the condition $\lambda_{min}(\alpha_c) = 0$ is a stability threshold (iteration procedure is unstable below the threshold and vice versa). The critical values of the iteration parameter straightforwardly follows from (5.7):

$$\alpha_c = \frac{\xi-1}{\xi+1}, \text{ where } \xi = \frac{\max(R_{0i})}{n} \quad (5.12)$$

Thus, the introduction of the iteration parameter is of essential importance, since the numerical simulations are unstable for $\alpha = 0$.

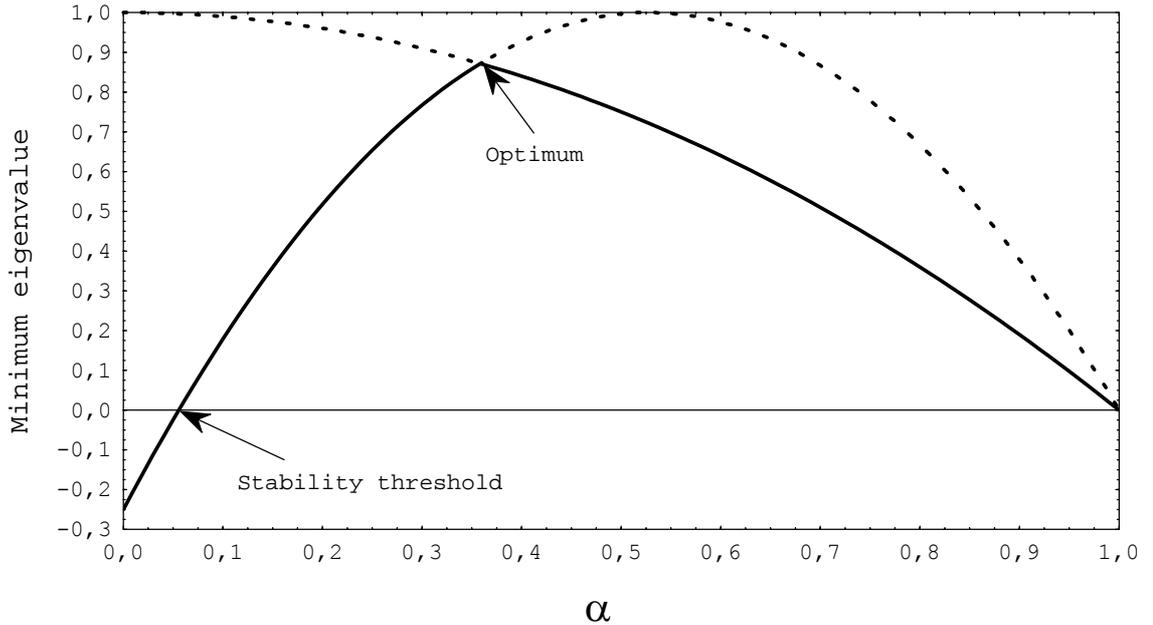

Fig 1. A dependence of the minimum eigenvalue of the contracting matrix on the iteration parametr

An example of the analysis of convergence of iteration procedure is shown in Fig. 1. The dependence $\lambda_{min}(\alpha)$ to be found is shown by the solid line formed by two segments of parabolas. Figure 1 corresponds to the data shown below in Fig. 2 (the sample of the size $n = 200$ from the mixture of two Gaussian components). Numerical simulations performed for various distributions (mixture of two and three Gaussian components, gamma distribution, beta distribution etc.) show that the dependence shown in the Fig.1 is universal (slightly changes with varying nature of data). In approximate estimations, one can assume that $\min(R_{0i}) \approx 0$ and $\max(R_{0i}) \approx n$; therefore, (5.8) and (5.7) result in following universal values $\alpha_{opt} \approx 1/3$ and $\lambda_{min}(\alpha_{opt}) \approx 1 - \alpha_{opt}^2 \approx 0{,}89$ for rough estimates. This means that at each iteration step, the square of the distance between the approximate and exact solutions (together with the difference of logarithmic likelihoods) decreases approximately by a factor of 9. In our example, the numerical simulations result in $\min(R_{0i}) \approx 0{,}0009662$ and $\max(R_{0i}) \approx 223{,}6$. The fact that the last number slightly exceeds the sample size ($n = 200$) yields the instability of iteration procedure at small values of the iteration parameter ($\alpha < \alpha_c$). In our case, the critical and optimal values are $\alpha_c \approx 0{,}0557$, $\alpha_{opt} \approx 0{,}359$, $\lambda_{\min}(\alpha_{opt}) \approx 0{,}871$, respectively.

## 6. Optimization of the Number of Harmonics

Let an exact (usually, unknown) psi function be

$$\psi(x) = \sum_{i=0}^{\infty} c_i \varphi_i(x). \quad (6.1)$$

Represent the psi-function estimator in the form

$$\hat{\psi}(x) = \sum_{i=0}^{s-1} \hat{c}_i \varphi_i(x). \quad (6.2)$$

Here, the statistical estimators are denoted by caps in order to distinguish them from exact quantities.

Comparison of two formulas shows that difference between the exact and estimated psi functions is caused by two reasons. First, we neglect $s$-th and higher harmonics by truncating the infinite series. Second, the estimated Fourier series coefficients (with caps) differ from unknown exact values.

Let $\quad \hat{c}_i = c_i + \delta c_i. \quad (6.3)$

Then, in view of the basis orthonormality, the squared deviation of the exact function from the approximate one may be written as

$$F(s) = \int (\psi - \hat{\psi})^2 dx = \sum_{i=0}^{s-1} \delta c_i^2 + \sum_{i=s}^{\infty} c_i^2. \quad (6.4)$$

Introduce the notation

$$Q(s) = \sum_{i=s}^{\infty} c_i^2. \quad (6.5)$$

By implication, $Q(s)$ is deterministic (not a random) variable. As for the first term, we will consider it as a random variable asymptotically related to the chi-square distribution:

$$\sum_{i=0}^{s-1} \delta c_i^2 \sim \frac{\chi_{s-1}^2}{4n},$$

where $\chi_{s-1}^2$ is the random variable with the chi-square distribution of $s-1$ degrees of freedom.

Thus, we find that $F(s)$ is a random variable of the form

$$F(s) = \frac{\chi_{s-1}^2}{4n} + Q(s). \quad (6.6)$$

We will look for an optimal number of harmonics using the condition for minimum of the function mean value $\overline{F}(s)$:

$$\overline{F}(s) = \frac{s-1}{4n} + Q(s) \to \min. \quad (6.7)$$

Assume that, at sufficiently large $s$,
$$Q(s) = \frac{f}{s^r}. \quad (6.8)$$

The optimal value resulting from the condition $\dfrac{\partial \overline{F}(s)}{\partial s} = 0$ is

$$s_{opt} = \sqrt[r+1]{4rfn}. \quad (6.9)$$

The formula (6.9) has a simple meaning: the Fourier series should be truncated when its coefficients become equal to or smaller than the error, i.e., $c_s^2 \leq \dfrac{1}{4n}$. From (6.8) it follows that $c_s^2 \approx \dfrac{rf}{s^{r+1}}$. The combination of the last two formulas yields the estimation (6.9).

The coefficients $f$ and $r$ can be calculated by the regression method. The regression function (6.8) is smoothed by taking a logarithm
$$\ln Q(s) = \ln f - r \ln s. \quad (6.10)$$

Another way to numerically minimize $F(s)$ is to detect the step when the inequality $c_s^2 \leq \dfrac{1}{4n}$ is systematically satisfied. It is necessary to determine that the inequality is met just systematically in contrast to the case when several coefficients are equal to zero in result of the symmetry of the function.

Our approach to estimate the number of expansion terms does not pretend to high rigor. For instance, strictly speaking, our estimation of the statistical noise level does not directly concern the coefficients in the infinite Fourier series. Indeed, the estimation was performed in the case when the estimated function is exactly described by a finite (preset) number of terms in the Fourier series with coefficients involving certain statistical error due to the finite size of a sample. Moreover, since the function to be determined is unknown (otherwise, there is no a problem), any estimation of the truncation error is approximate, since it is related to the introduction of additional assumptions.

The optimization of the number of terms in the Fourier series may be performed by the Tikhonov regularization methods [20].

### 7. Numerical Simulations. Chebyshev-Hermite Basis

In this paper, the set of the Chebyshev - Hermite functions corresponding to the stationary states of a quantum harmonic oscillator is used for numerical simulations. In particular, this basis is convenient since the Gaussian distribution in zero-order approximation can be achieved by choosing the ground oscillator state; and adding the contributions of higher harmonics into the state vector provides deviations from the gaussianity.

The set of the Chebyshev- Hermite basis functions is [23]

$$\varphi_k(x) = \frac{1}{\left(2^k k! \sqrt{\pi}\right)^{1/2}} H_k(x) \exp\left(-\frac{x^2}{2}\right), \quad k = 0,1,2,\ldots. \quad (7.1)$$

Here, $H_k(x)$ is the Chebyshev- Hermite polynomial of the $k$-th order. The first two polynomials have the form
$$H_0(x) = 1, \quad (7.2)$$
$$H_1(x) = 2x. \quad (7.3)$$
The other polynomials can be found by the following recurrent relationship:

$$H_{k+1}(x) - 2xH_k(x) + 2kH_{k-1}(x) = 0. \quad (7.4)$$

The algorithms proposed here have been tested by the Monte Carlo method using the Chebyshev-Hermite functions for a wide range of distributions (mixture of several Gaussian components, gamma distribution, beta distribution etc.). The results of numerical simulations show that the estimation of the number of terms in the Fourier series is close to optimal. It turns out that the approximation accuracy decays more sharply in the case when less terms than optimal are taken into account than in the opposite case when several extra noise harmonics are allowed for. From the aforesaid, it follows that choosing larger number of terms does not result in sharp deterioration of the approximation results. For example, the approximate density of the mixture of two components weakly varies in the range from 8—10 to 50 and more terms for a sample of several hundreds points.

The quality of approximation may be characterized by the following discrepancy between the exact density and its estimation ($L_1$ - norm) [24]: $\Delta = \int |\hat{p}(x) - p(x)| dx$. The properties of a random variable $\Delta$ can be studied by the Monte Carlo method. Several results are listed in Table 1.

| Table 1. | | | |
|---|---|---|---|
| Estimator | $\mu_1 = 0$  $\mu_2 = 3$  $f_1 = 0{,}7$  $f_2 = 0{,}3$ | $\mu_1 = 0$  $\mu_2 = 3$  $f_1 = f_2 = 0{,}5$ | $\mu_1 = 0$  $\mu_2 = 3$  $\mu_3 = -3$  $f_1 = 0{,}5$  $f_2 = f_3 = 0{,}25$ |
| Root | 0,0978  (0,033) | 0,106  (0,035) | 0,119  (0,030) |
| Kernel | 0,136  (0,032) | 0,143  (0,034) | 0,147  (0,027) |
| Orthogonal series | 0,151  (0,043) | 0,147  (0,041) | 0,169  (0,038) |

In Table 1, mean values and standard deviations (in parentheses) for a random variable $\Delta$ obtained by the Monte Carlo method are listed for root, kernel (Rosenblatt - Parzen) [25-27], and orthogonal series (Chentsov) [28-31] estimators in problems of density estimations for the mixtures of two and three Gaussian components. The mathematical expectations and weights of components are shown in table. The standard deviation is equal to unity for each component. The sample size is 200; 100 trials are made in each numerical experiment. The results of this study and other similar researches reliably indicate the essential advantages of the root estimator compared to the other.

Figure 2 (upper plot) exemplifies the comparison between the root density estimator, the Rosenblatt - Parzen kernel density estimator, and orthogonal series estimator proposed by Chentsov. Nine terms are taken into account in expansions of the psi function and density for both the root and orthogonal series estimators in the Chebyshev-Hermite basis. The bandwidth of Gaussian kernel density is calculated as a sample standard deviation divided by the 5th root of the sample size.

The bottom plot in Fig. 2 illustrates the rectification of the dependence according to (6.10) (the sample size is $n = 200$). As is noted above, Figs. 1 and 2 correspond to the same statistical data.

This approach implies that the basis for the psi-function expansion can be arbitrary but it should be preset. In this case, the found results turn out to be universal (independent of basis). This concerns the Fisher information matrix, covariance matrix, chi-square parameter etc. The set of the Chebyshev-Hermite functions can certainly be generalized by introducing translation and scaling parameters that have to be estimated by the maximum likelihood method. The Fisher information matrix, covariance matrix etc. found in such a way would be related only to the Chebyshev-Hermite basis and nothing else.

Practically, the expansion basis can be fixed beforehand if the data describe a well-known physical system (e.g., in atomic systems, the basis is preset by nature in the form of the set of stationary states).

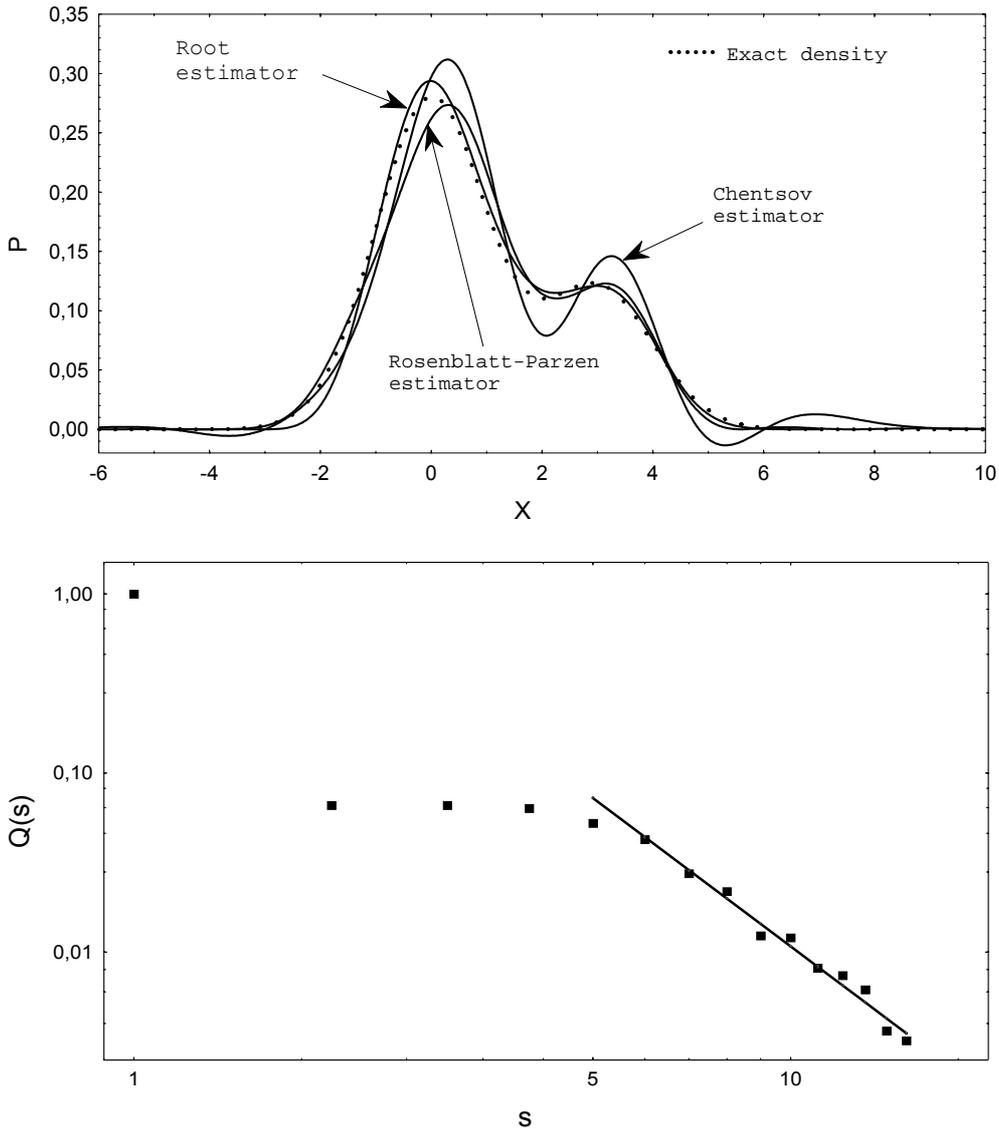

Fig.2 (a) An example of comparison of the Root estimator with the Rosenblatt-Paezen and Chentsov estimators; (b) smoothing the dependence by (6.10)

In other cases, the basis has to be chosen in view of the data under consideration. For instance, in the case of the Chebyshev-Hermite functions, it can be easily done if one assumes that the distribution is Gaussian in the zero-order approximation.

Note that the formalism presented here is equally applicable to both one-dimensional and multidimensional data. In the latter case, if the Chebyshev-Hermite functions are used, one may assume that multidimensional normal distribution takes place in the zero-order approximation that, in its turn, can be transformed to the standard form by translation, scale, and rotational transformations.

## 8. Density Matrix

The density matrix method is a general quantum mechanical method to study inhomogeneous statistical populations (mixtures) [5, 32]. The corresponding technique can be used in statistical data analysis as well.

First, for example, consider a case when the analysis of Sec.2 shows that two statistical samples are inhomogeneous. In this case, the density matrix represented by a mixture of two components can be constructed for the total population:

$$\rho = \frac{n_1}{n}\rho^{(1)} + \frac{n_2}{n}\rho^{(2)}, \quad (8.1)$$

where $n = n_1 + n_2$,

$$\rho_{ij}^{(1)} = c_i^{(1)} c_j^{(1)*}, \quad (8.2)$$

$$\rho_{ij}^{(2)} = c_i^{(2)} c_j^{(2)*}. \quad (8.3)$$

Any density matrix can be transformed to the diagonal form by a unitary transformation. In the diagonal representation, the density matrix of the pure state will have the only element equal to unity and the other, equal to zero. In the case of two component mixture (8.1), there will be two nonzero elements etc.

Note that only the density matrix of a pure state satisfies the condition

$$\rho^2 = \rho. \quad (8.4)$$

In the diagonal representation, the density for the mixture of $m$ components can be represented in terms of eigenvalues and eigenfunctions of the density matrix:

$$p(x) = \sum_{i=1}^{m} \lambda_i |\psi_i(x)|^2. \quad (8.5)$$

In the case when the first component prevails in the expansion (8.5), it may be considered as responsible for the basic density; whereas the other, describing noise.

Now, we cite some information on the density matrix from quantum mechanics.

The mean value of a physical quantity $A$ is expressed in terms of psi function as follows:

$$\overline{A} = \int \psi^*(x) A(x) \psi(x) dx =$$
$$= c_i c_j^* \int \varphi_j^*(x) A(x) \varphi_i(x) dx = \rho_{ij} A_{ji} = Tr(\rho A). \quad (8.6)$$

Here, the density matrix $\rho$ and the matrix element of $A$ are given by

$$\rho_{ij} = c_i c_j^*, \quad (8.7)$$

$$A_{ji} = \int \varphi_j^*(x) A(x) \varphi_i(x) dx, \quad (8.8)$$

and $Tr$ denotes the trace of a matrix.

The density matrix introduced in such a way relates to a so-called pure state described by a psi function. In general case, a matrix satisfying the following three conditions can be referred to as a density matrix:
1. Hermitian matrix:

$$\rho^+ = \rho. \quad (8.9)$$

2. Positive (nonnegative) matrix

$$\langle z|\rho|z\rangle \equiv \sum_{i,j} \rho_{ij} z_i^* z_j \geq 0 \quad (8.10)$$

for any column vector $|z\rangle$. The sign of equality takes place only for the identically zero column vector. Thus the diagonal elements of the density matrix are always nonnegative.
3. The matrix trace is equal to unity:

$$Tr(\rho) = 1 . \quad (8.11)$$

Each density matrix $\rho$ defined on a orthonormal basis $\varphi_i(x)$ may be placed in correspondence with a density operator

$$\rho(x, x_1) = \rho_{ij}\varphi_i(x_1)\varphi_j^*(x) . \quad (8.12)$$

In the case when the arguments of the density operator coincide, we obtain the basic object of probability theory, namely, probability density

$$p(x) = \rho(x, x) = \rho_{ij}\varphi_i(x)\varphi_j^*(x) . \quad (8.13)$$

The only probability density corresponds to the density matrix (in a given basis). The opposite statement is incorrect.

The mean value (mathematical expectation) of any physical quantity given by an arbitrary operator $A$ is

$$\overline{A} = \rho_{ij}A_{ji} = Tr(\rho A) . \quad (8.14)$$

Now, consider the case when the analysis by the algorithm of Sec.2 shows the homogeneity of both statistical samples. In this case, the state estimator for the total population should be represented by a certain superposition of the state estimators for separate samples. Let us show that the corresponding optimal estimator is the first principle component of the joint density matrix (8.1) in the expansion (8.5).

Let us express the eigenvectors of the joint density matrix (8.1) in terms of eigenvectors of the components:

$$\rho_{ij}c_j = \lambda c_i , \quad (8.15)$$

$$c_j = a_1 c_j^{(1)} + a_2 c_j^{(2)} . \quad (8.16)$$

Substituting (8.16) into (8.15), in view of (8.2) and (8.3), we have a set of the homogeneous equations in two unknowns:

$$\left.\begin{array}{l}(n_1 - \lambda(n_1 + n_2))a_1 + n_1 r a_2 = 0 \\ n_2 r^* a_1 + (n_2 - \lambda(n_1 + n_2))a_2 = 0\end{array}\right\}, \quad (8.17)$$

where

$$r = (c^{*(1)}, c^{(2)}) = c_i^{*(1)} c_i^{(2)} \quad (8.18)$$

is a scalar product of two vectors.

The system admits a solution if its determinant is equal to zero. Finally, the eigenvalues of the joint density matrix are

$$\lambda_{1,2} = \frac{1 \pm \sqrt{1 - 4k}}{2} , \quad (8.19)$$

where

$$k = \frac{n_1 n_2 (1 - |r|^2)}{(n_1 + n_2)^2} . \quad (8.20)$$

For simplicity's sake, we restrict our consideration to real valued vectors (just as in Sec.2). In the case of homogeneous samples, asymptotically $1 - r^2 = (1 + r)(1 - r) \to 2(1 - r)$. Then, according to (2.5)-(2.6), we asymptotically have

$$4k = \frac{\chi^2_{s-1}}{n_1+n_2} \sim O\left(\frac{1}{n_1+n_2}\right). \quad (8.21)$$

Then, $\lambda_1 = 1 - O\left(\dfrac{1}{n_1+n_2}\right)$, (8.22)

$$\lambda_2 = O\left(\frac{1}{n_1+n_2}\right). \quad (8.23)$$

Thus, the first principal component has maximum weight while merging two homogeneous samples. The second component has a weight of an order of $\dfrac{1}{n_1+n_2}$ and should be interpreted as a statistical fluctuation. If one drops the second component, the density matrix would become pure.

Equation (8.17) and the normalization condition yield

$$a_1 \approx \frac{n_1}{n_1+n_2} \quad a_2 \approx \frac{n_2}{n_1+n_2} \quad (8.24)$$

up to terms of an order of $\dfrac{1}{n_1+n_2}$. In view of (1.19), the deviation of the resulting state vector is

$$\xi = a_1 \xi^{(1)} + a_2 \xi^{(2)}, \quad (8.25)$$

$$E(\xi_i \xi_j) = a_1^2 E(\xi_i^{(1)} \xi_j^{(1)}) + a_2^2 E(\xi_i^{(2)} \xi_j^{(2)}) =$$

$$= \frac{1}{4(n_1+n_2)}(\delta_{ij} - c_i c_j) \quad i,j = 0,1,...,s-1. \quad (8.26)$$

The last equation shows that the fist principal component of a joint density matrix is asymptotically efficient estimator of an unknown state vector.

Thus, in merging two large homogeneous samples with certain state estimators, it is not necessary to return to the initial data and solve the likelihood equation for a total population. It is sufficient to find the first principle component of a joint density matrix. This component will be an estimator for the state vector of statistical ensemble that is refined over the sample population. As it has been shown above, such an estimator is asymptotically effective and is not worth than the estimator on the basis of initial data (to be precise, the loss in accuracy is of a higher order of magnitude than (8.26)).

This property is of essential importance. It implies that the estimated psi function involves practically all useful information contained in a large sample. In other words, psi function is asymptotically sufficient statistics. This property can also be interpreted as an asymptotic quasilinearity of a state vector satisfying the nonlinear likelihood equation.

**Conclusions**

Let us state a short summary.

Search for multiparametric statistical model providing stable estimation of parameters on the basis of observed data results in constructing the root density estimator. The root density estimator is based on the representation of the probability density as a squared absolute value of a certain function, which is referred to as a psi function in analogy with quantum mechanics. The method proposed is an efficient tool to solve the basic problem of statistical data analysis, i.e., estimation of distribution density on the basis of experimental data.

The coefficients of the psi-function expansion in terms of orthonormal set of functions are estimated by the maximum likelihood method providing optimal asymptotic properties of the method (asymptotic unbiasedness, consistency, and asymptotic efficiency). An optimal number of harmonics in the expansion is appropriate to choose, on the basis of the compromise, between two opposite tendencies: the accuracy of the estimation of the function approximated by a finite series increases with increasing number of harmonics, however, the statistical noise level also increases.

The likelihood equation in the root density estimator method has a simple quasilinear structure and admits developing an effective fast-converging iteration procedure even in the case of multiparametric problems. It is shown that an optimal value of the iteration parameter should be found by the maximin strategy. The numerical implementation of the proposed algorithm is considered by the use of the set of Chebyshev-Hermite functions as a basis set of functions.

The introduction of the psi function allows one to represent the Fisher information matrix as well as statistical properties of the sate vector estimator in simple analytical forms. Basic objects of the theory (state vectors, information and covariance matrices etc.) become simple geometrical objects in the Hilbert space that are invariant with respect to unitary (orthogonal) transformations.

A new statistical characteristic, a confidence cone, is introduced instead of a standard confidence interval. The chi-square test is considered to test the hypotheses that the estimated vector equals to the state vector of general population and that both samples are homogeneous.

It is shown that it is convenient to analyze the sample populations (both homogeneous and inhomogeneous) using the density matrix.

**Appendix 1.**
**Maximum Likelihood Method and Fisher Information Matrix**

Let $x = (x_1,...,x_n)$ be a sample under consideration represented by $n$ independent observations from the same distribution $p(x|\theta)$. Here, $\theta$ is the distribution parameter (in general, vector valued).

The likelihood function is determined by the following product:

$$L(x|\theta) = \prod_{i=1}^{n} p(x_i|\theta) \, . \text{ (A1.1)}$$

The formula under consideration is the $n$-dimensional joint density of random distributions of the components of the vector $x = (x_1,...,x_n)$ interpreted as a set of independent random variables with the same distribution. But if $x = (x_1,...,x_n)$ is a certain realization (fixed sample), the likelihood function as a function of $\theta$ characterizes the likeliness of various values of the distribution parameter.

According to the maximum likelihood principle put forward by Fisher in 1912 [33] and developed in the twenties of the last century [34], the value $\hat{\theta}$ from the region of acceptability, where the likelihood function reaches its maximum value, should be taken as an estimation for $\theta$.

As a rule, it is more convenient to deal with the log likelihood function:

$$\ln L = \sum_{i=1}^{n} \ln p(x_i|\theta). \quad (A1.2)$$

Both the log likelihood and likelihood functions have extrema at the same points due to the monotonicity of the logarithm function.

The necessary condition for the extremum of the log likelihood function is determined by the likelihood equation of the form

$$\frac{\partial \ln L}{\partial \theta} = 0. \quad (A1.3)$$

If $\theta = (\theta_1, ..., \theta_s)$ is an $s$-dimensional parameter vector, we have the set of the likelihood equations

$$\frac{\partial \ln L}{\partial \theta_i} = 0 \quad i = 1, ..., s \quad (A1.4)$$

The basic result of the theory of maximum likelihood estimation is that under certain sufficiently general conditions, the likelihood equations have a solution $\hat{\theta} = (\hat{\theta}_1, ..., \hat{\theta}_s)$ that is a consistent, asymptotically normal, and asymptotically efficient estimator of the parameter $\theta = (\theta_1, ..., \theta_s)$ [12-14].

Formally, the aforesaid may be expressed as

$$\hat{\theta} \sim N(\theta, I^{-1}(\theta)). \quad (A1.5)$$

The last formula means that the estimator $\hat{\theta}$ is asymptotically (at large $n$) a random variable with a multidimensional normal distribution with the mean equal to the true value of the parameter $\theta$ and the covariance matrix equal to the inverse of the Fisher information matrix.

The Fisher information matrix elements are

$$I_{ij}(\theta) = n \cdot \int \frac{\partial \ln p(x|\theta)}{\partial \theta_i} \frac{\partial \ln p(x|\theta)}{\partial \theta_j} p(x|\theta) dx \quad (A1.6)$$

The factor $n$ indicates that the Fisher information is additive (the information of a sample consists of the information in its points). At $n \to \infty$, the covariance matrix asymptotically tends to zero matrix, and, in particular, the variances of all the components become zero (consistency).

The fundamental significance of the Fisher information consists in its property to set the constraint on achievable (in principle) accuracy of statistical estimators. According to the Cramer-Rao inequality [12-14], the matrix $\Sigma(\hat{\theta}) - I^{-1}(\theta)$ is nonnegative for any unbiased estimator $\hat{\theta}$ of an unknown vector valued parameter $\theta$. Here, $\Sigma(\hat{\theta})$ is the covariance matrix for the estimator $\hat{\theta}$. The corresponding difference asymptotically tends to a zero matrix for the maximum likelihood estimators (asymptotic efficiency).